\documentclass[12pt,a4paper]{article}
\usepackage{amsmath}
\usepackage{amsfonts}
\usepackage{amssymb}

\author{Andrei Khrennikov\\
International Center for Mathematical Modeling \\in Physics, Engineering, Economics, and Cognitive Science \\
Linnaeus University, S-35195, V\"axj\"o, Sweden}

\title{Unuploaded experiments have no result}

\begin{document}

\maketitle

\abstract{The aim of this note is to attract once again attention of the quantum community to statistical analysis of data which was reported as violating 
Bell's inequality. This analysis suffers of a number of problems. And the main problem is that rough data is practically unavailable. 
However, experiments which are not followed by the open access to the rough data have to be considered 
as with no result. The absence of rough data generates a variety of problems in statistical interpretation of the results of Bell's type experiment. 
One may hope that this
 note would stimulate experimenters to create the open access data-base for, e.g.,  Bell tests.  Unfortunately, recently announced 
experimental loophole-free violation of a Bell inequality using entangled electron spins separated by 1.3 km was not supported by  open-access data.
Therefore in accordance with our approach ``it has no result.'' The promising data after publication is, of course, a step towards fair analysis quantum experiments. 
May be this is a consequence of appearance of this preprint, v1. But there are  a few questions  which would be 
interesting to clarify before publication (and which we shall discuss in this note).}

\section{Introduction}

By publishing the previous version of this preprint I hoped that it would have some impact to experimenters performing quantum foundational experiments and first of all experiments on 
 violation of the Bell inequality. For the latter  problem, availability of rough data for independent statistical analysis is very important. During the last years a few leading experimental groups demonstrated
that they are very close to close all possible (or at least basic) loopholes in the Bell test, we mention, e.g., Zeilinger's group \cite{Z} and Kwiat's group \cite{K}. The rough data for such ``finalizing 
experiments'' is till not approachble. Here the main problem is the statistical significance of the announced results, see \cite{ZH}--\cite{Vienna} for discussions. 

Without rough data, it may happen that there will be declared that the loophole free test has been successfully performed. After this further experimental studies would be 
considered as meaningless and the result which may be not  justified from the viewpoint of an independent statistical expertise would be everywhere cited as the final result 
closing the Bell-violation problem. 

It seems that v1 of this preprint had some impact to experimenters. For example,  the team announced 
{\it ``experimental loophole-free violation of a Bell inequality using entangled electron spins separated by 1.3 km''}
 \cite{i1} promised open-access to rough data after publication. This is definitely  
an important step  towards fair analysis of quantum experiments. 
However, there are a few questions which would be interesting to clarify even before publication. 
We shall mention some of them in new section, section \ref{N}, completing v1 of this preprint. Thus the reader who have already read v1 can jump directly to section \ref{N}.

\medskip

One of the basic principles of the scientific methodology is that experimental data have to be approachable by researchers independent from the group performed 
a reported experiment. Statistics is a very delicate area of research and it is worse to provide a possibility to check statistical conclusions which are typically 
presented in the compact form in articles of experimenters (PhD-thesis  are better, but there are still no rough click by click data, although nowadays data can be easily 
uploaded to the corresponding website containing the pdf-file of the PhD-thesis). Unfortunately, to put data on lab's  website 
is not a tradition in the quantum community, opposite to others, e.g.,   
molecular biology and genetics as well as psychology. This situation has to be improved, as soon as possible. A. Peres famously pointed out that {\it ``Unperformed 
experiments have no results''} \cite{AP}; I would add that {\it ``Unuploaded experiments have no result.''}

Nowadays experimental testing of violations of Bell's inequality \cite{B0}, \cite{B} is one of the most important foundational projects. It also plays the determining role 
for some quantum technologies, especially quantum random generators \cite{Pironio}\footnote{If I were planing to buy a quantum random generator, I would definitely like to check claims of sellers 
on the real experimental data. Of course, at the moment this problem is not actual, because we are still very far from the loophole free test for violation of the 
Bell inequality.}. Therefore impossibility of open access to click by click data for such tests is really a problem. 
Quantum foundational community really needs such data. However, even data from other quantum foundational experiments (and not only with photons)  is badly demanded. 

In spite of my very wide network of contacts with experimenters, I was  able to approach only   one set of rough data, click by click, 
from the experiment of G. Weihs \cite{Weihs0}, \cite{Weihs}.
However, this experiment was done long ago. Since that time the quantum experimental technology was improved crucially. 
In particular, at that time there were used detectors of low efficiency and it 
might be that majority of ``strangenesses in Weihs' data'' are simply 
consequences of the low detection efficiency. Another my source of data is even older, this is the PhD-thesis of A. Aspect \cite{ Aspect0}. 
These data suffers of the same problems which might be again simply consequences  of low efficiency of  detectors.  

The absence of data or even worse consequences of the statistical analysis of the available data  \cite{Adenier} 
in combination with impossibility to get new data 
lead to a number of problems which permanently disturb my mind  and make me really suffering. 
Among them I would like to mention the following four problems which will be discussed in the section 2:  
\begin{itemize}
\item statistical justification of  no-signaling; 
\item ``Aspect's type anomalies'' in   data; 
\item the time window problem (coincidence loophole);
\item statistical analysis of Bell's experiment.    
\end{itemize}

We also remark that the length of Bell's experimenting and the strong wish to consider this problem as ``practically closed'' made the  quantum community  tiered; 
majority of the community lives with the very strong belief that the problem was completely clarified -- depending on a person it can be said  that it was done 
 ``already by Aspect'' or ``by Weihs'',  ``by Zeilinger's team'', by ``Kwiat's group''. Yes, the sudy is much longer than one could expect at the beginning, but it is still far from to be completed.
And the problem is not only to close all loopholes (as many naively believe), but to perform a proper statistical analysis of data for experiments declaring 
closing of particular loopholes.   For a moment, I can mention only three papers which authors attempted to perform such an analysis, \cite{Adenier}, \cite{Pironio}, 
\cite{ZH}. 

After the recent experiments closing the detection loophole\footnote{These were really great experiments closing one of the 
most important loopholes (in my opinion the most important) and I propose to assign them the names, e.g., by the places were they have been done.}, 
the Vienna experiment \cite{Z} and the Urbana-Champaign experiment\cite{K}, it is commonly accepted that there is only one step to complete Bell's project: to close
in a single experiment both locality and detection efficiency loopholes. I agree that such an experiment will of the great value, both for quantum foundations and technologies.
However, it would be still not the end of the Bell-story. It has to be completed by the detailed statistical analysis of data and I expect that it will take a long time to approach 
consensus on the results of this analysis. I repeat again that statistics is a delicate science.

One can imagine a kind of EU project on ``certification'' of violation of Bell's type inequalities: each experiment closes the concrete loophole or a group of loopholes 
and the complete data are in open access. It would be not so bad to double such a data-base by certifying violations of Bell's inequality by some leading group, e.g., in USA. 
Such certification is really needed if  we plan to proceed to quantum technologies, especially based on quantum random generators and (less) on quantum cryptography. 
Moreover, the majority of the basic quantum foundational experiments have been done long ago and the rough data is neither available for the open access. Where can one find data 
for the two slit experiment with photons? with electrons? data of Weihs' three slit experiment \cite{Weihs1}? experiments of neutron interferometry? 
experiments on ``photon existence'' -- estimation of the coefficient of second order coherence $g^{2}(0)$ (with heralded photons and with a geniune single photon source)?

Besides to secure the users of the quantum technologies that in reality everything matches theory,   such a data-base would simplify essentially the foundational debates and 
restrict the opposition to the conventional interpretation of violation of Bell's inequality: only conclusions of people educated in statistics would be considered seriously.
Nowadays the absence of the rough data provides possibilities for practically everybody to speculate and say: ``well, there is no data, it might be that, in fact, ...''

\section{Three issues disturbing me}

\subsection{Signaling}

My impression is that experimental groups reporting a violation of  Bell's inequality  do not check the condition of {\it no-signaling.}
 They are fine by violating the Bell inequality by as many $\sigma$ as they can. At the same time,  the data which I was able to get always violate
the condition of no-signaling. Of course, the main problem was  that these data were really old. And there are practically 
no possibility to get new rough data. 

I start with the remark that, in fact, {\it the terminology no-signaling is  ambiguous.} 
In reality we want just to check that the marginal probabilities on side $S_1$ obtained by summation with respect to the results on the other 
side $S_2$ do not depend on settings on $S_2:$ 
\begin{equation}
\label{NS}
p_{i; L}(+) = p_{ij}(+ +) + p_{ij}(+ -), \; p_{i;L}(-) = p_{ij}(- +) + p_{ij}(- -),  
\end{equation}
where $i,j$ encode experimental settings, the angles $\theta_i, \theta_j$ of the orientation of polarization beam 
splitters at the left-hand and  right-hand sides labs. In the same way
\begin{equation}
\label{NS1}
p_{j;R}(+) = p_{ij}(+ +) + p_{ij}(- +), \; p_{j;R}(-) = p_{ij}(+ -) + p_{ij}(- -). 
\end{equation}
 This is one of necessary conditions for existence of the classical probability distribution serving for the experiment, see, e.g.,  \cite{CONT}, \cite{Brukner} 
for details. This is simply the condition of additivity of probability.\footnote{Violation of this condition in the two slit experiment, where one of indexes is used to label 
a slit, and another to label a point at the registration photo-emulsion screen, was discussed by Feynman \cite{Feynman}. The general approach to interference
as violation of additivity of probability was elaborated in a series of my works, e.g., \cite{CONT}. } However, by following the physical tradition we shall keep to the terminology (no-)signaling.

The presence of ``signaling'' in experimental data  is very disturbing for me. In Aspect's experiment \cite{Aspect}, the assumption of no-signaling is violated (this can be extracted
from his PhD thesis \cite{Aspect0}.) This assumption is violated by Weihs' data as well \cite{Adenier}. We shall discuss later  possible sources of violation 
of no-signaling.

 Now I want to make a point \cite{Adenier1}:

\medskip

 Formally, experimental data violating both Bell's inequality and no-signaling either cannot be used against 
local realism or such data have to be used to argue that both local both local realism and QM have to be rejected.  The latter predicts no-signaling. You got signaling? 
Then you have to reject QM. One of, course, does not want to proceed in this way and experimenter would come with detailed explanation of all 
technicalities which lead to signaling. I do not question these explanations; experts know technicalities perfectly. So, the role of these technicalities have to  be taken seriously.
However, in this situation it is logically reasonable to accept that violations of Bell's inequality might be explained by other technicalities and take seriously the arguments of 
such a type, e.g.,  \cite{Entropy}, \cite{Hans}, \cite{Marian}.

\medskip

Even by having data one confronts another extremely difficult problem, namely, the problem of selecting of proper statistical test for no-signaling.   
Perhaps the most complete analysis of a Bell-type experiment is the
one given by Pironio et al. \cite{Pironio}.  They performed
statistical tests to check for signaling problems (as we defined it as consistency condition for marginal distributions),
see the supplementary information of this paper. However, I cannot point to other publications on this problem and  
of such a level of statistical analysis. At the same time applicability of  the statical test used in  \cite{Pironio}
to data from Bell's type experiments can be questioned, and it was questioned in   Yanbao Zhang et al. \cite{ZH}, 
where a detailed analysis of specialties of such data.  Therefore it is important to have open access data, click by click, which can be used 
by experts in statistics. 

\subsection{``Anomalies'' in Aspect's  data}

A few years ago  Alain Aspect told me about some strange anomalies which he founded in his data, 
data from the pioneer experiment \cite{Aspect0},  and these anomalies can be found in his thesis. 
That his data exhibited a strange behavior. If we take the joint probabilities 
\begin{equation}
\label{LL}
p_{ij}(+,+),..., p_{ij}(-,-),
\end{equation}
where $i,j$ encode experimental settings, then these probabilities
differ  from the prediction of QM for maximally entangled state,
i.e., from 
\begin{equation}
\label{LL1}
p_{ij}(+ +)= p_{ij}(- -)=
1/2 \cos^2(\theta_i- \theta_j)/2,   
\end{equation}
\begin{equation}
\label{LL2}
p_{ij}(+ -)= p_{ij}(+ -)=  1/2 \sin^2(\theta_i- \theta_j)/2. 
\end{equation}
However, at the same time the expressions for correlations
\begin{equation}
\label{LL3}
E(ij) =p_{ij}(+ +) - p_{ij}(+ -) - p_{ij}(- +)+ p_ij(- -)
\end{equation}
match very well with ones calculated from theory, for the maximally entangled state. 
Probabilities in  $E(ij)$ compensate each other in a mysterious way. The same we have seen in Weihs' data \cite{Adenier}.
Our attempt to solve the latter problem by considering non-maximally entangled states was not successful (calculations and simulation were performed
by Adenier \cite{Adenier}); we were neither satisfied by Weihs' explanation in terms of mixed states \cite{Weihs2}. 

I have a plenty of discussions about these anomalies with the top experts; they pointed to a variety of technicalities which might 
lead to the anomalies. The main issue is that it might be that these {\it  experiments were calibrated for correlations and not for states}.
The experimenters also pointed to possible instabilities in the state production, pair production rates and measurement settings. I suppose
that in modern experiments these characteristics were essentially improved. However, I am not completely sure, because by 
attacking new loopholes experimenters met tremendous new challenges and they often ignore smallnesses as, e.g., stability 
of the state production or the pair rate production. 

For an expert in quantum foundation who is not an experimenter by origin, the statement   
that {\small these experiments were calibrated for correlations and not for states} is really disturbing.
Would it be in principle possible to ``calibrate'' an experiment in such a way that all loopholes
including such ``smallnesses''  as no-signaling and the absence of the Aspect-type anomalies be combined?
And not simply combined, but  with a sufficiently high level of statistical confidence for all of them?
 Do there exist {\it complimentary calibrations}?\footnote{Theoretical analysis of this type of problems was performed
 by the author and Volovich \cite{KHR_V}; we came to the conclusion that some loopholes are complementary and 
they would be never closed jointly. May be we were wrong in our theoretical analysis, but the modern experimental 
situation seems to be supporting our claim.}

In any event, we are still far from an experiment which will not suffer of aforementioned technicalities which can be in short 
characterized as the problem of the {\it stability of the experimental context} in long run quantum experiments.  

\subsection{The time window problem (coincidence loophole)}

I start again with a remark about the terminology. Although the terminology the ``coincidence
loophole'' is standard, I would prefer to speak about {\it the time window problem.} The latter rightly 
emphasized the center role of an experimenter playing with the size of the time window in correlation 
experiments. Moreover, there are some tinny differences in treating  the time-issue in the correlation 
experiments from the click-coincidence viewpoint and from time window selection viewpoint, see later discussion.

Regarding this problem, first of all  I  point to the contribution 
K. Hess and W. Philipp \cite{HP1}, \cite{HP2} who actively emphasized the role of time in the Bell experiment, as an additional 
hidden (``forgotten'') variable.\footnote{Their views matched well with my attempts to use the frequency von Mises theory to model violation 
of Bell's inequality \cite{KK}.} Their lovely debates during the V\"axj\"o series of conferences, first of all, with R. Gill and J. A. Larsson \cite{Gill},
\cite{Gill1} 
contributed a lot to clarification of the role of the coincidence/time window loophole. (Here I just emphasize the role of the debate,
without to comment the validity of the arguments of the both sides and their mathematical constructions).

Typically experts point out that the pioneer experiments of Aspect and Weihs suffer of the coincidence loophole. For the latter 
experiment, 

{\small ``Coincidences were identified by calculating time differences between Alice's and Bob's time tags and comparing these with a time window 
(typically a few ns).''}

To avoid the coincidence loophole, one do not consider differences
between time tags, but check the arrival time locally against an
absolute trial time window referenced to when the settings choice was
made.

Here we can see the difference between treating the time problem 
from the viewpoints of  time coincidence and time window. For the latter, 
coincidence-identification procedure does not play any role; the crucial point 
is the presence of selection of time window. This is especially clear from 
the analysis of Weihs' data performed by De Raedt et al. \cite{Hans} who claimed 
that the selection of the time window can be treated as a post-selection procedure,
that precisely by playing with the size of time window one violates Bell's inequality.
This is a very strong and even offensive statement. Therefore it is of the great interest
to be able to check its validity for new sets of data.  

\section{Statistical analysis of Bell's experiment}

\subsection{How many standard deviations?}

In Bell's experimenting essential efforts have been put to approach a violation of Bell's inequality with so many 
standard deviations as possible. Is such an activity really meaningful? Should an experimenter be proud by approaching 
such a result? My private opinion (shared with a few my friends-statisticians) is that  it is not so much meaningful to try 
to get more than  3-4 standard deviations. Simply mathematics says us that by increasing the number of trials $N$ 
we {\it automatically} increase the wanted number of $\sigma.$ Roughly speaking, this is just a measure of patience of 
a PhD-student collecting data, how many hours she can spend in the lab. 

\subsection{Independence of trials, stability of experimental conditions}

Of course, as everyone knows from the basics of probability theory, from the central limit theorem,  the number of  standard deviations can be used to 
measure statistical significance of violation of Bell's inequality only for a sequence of independent and equally distributed trials. If data contains
some dependence-patterns (and this often happens in real experiments\footnote{And if the rough data were public-available, we would see 
even more such dependence-patterns.}), then statistical significance cannot be simply reduced to a number of standard deviations. 

One has to use other methods to estimate a confidence interval and one of the simplest approaches (although rough enough) is to use Chebyshov inequality
as was done, e.g., in \cite{Vienna}; more advanced approaches were discussed in \cite{ZH}.  Thus the presence of dependences in data is not a big problem from the 
statistical viewpoint. Therefore may be for experimenters it is easier to apply more advanced statistical methods for data analysis, than to struggle 
with dependence-technicalities in the experiment.    
   
   \section{Experimental loophole-free violation of a Bell inequality using entangled electron spins separated by 1.3 km}
\label{N}

We turn to the recent experimental test announced in \cite{i1}. This is really an important step towards the final loophole free experiment, but the statistical significance of the result is questionable.

The biggest weakness is that {\bf the rate with which they perform their measurements is extremely low.}
 With approximately one event per hour (!), it takes them about 9 days to record a total of 245 coincidences (``trials of the Bell test''). 
Even the first experiment with atomic cascades performed in the 1970s (before Aspect's) had rates of pair productions that were still at least 100 times higher than this one.

Here just one CHSH measurement was made, and estimate the statistical significance of the violation is based on extra assumptions 
(independence, stability, normal distribution). Are these assumptions justified? Perhaps, but {\it extra assumptions are of course loopholes. }
The estimation of the standard deviation itself could be underestimated, and when the violation is just two standard deviation away from the absence of violation, 
{\bf this is not a minor issue!}

Quite simply, with such a low rate of detection, the experimental conditions are necessarily going to be different from one detection to the other. 
The authors  don't mention drifts at all in their paper, but there has to be some over the course of nine days and nights. 
Did they correct for those drifts? If not, then the assumption of normal distribution can be questioned, and if they did, well, what good 
is a violation of Bell inequality in which the experimenter is allowed to recalibrate/adjust/tune his experimental setup between each detection? 
Okay, the fact that they perform random measurements is in principle enough to protect them against such argument,
 but combined with the low number of counts, it remains quite problematic.

{\bf Another issue is no-signaling.} As was emphasized in previous sections, signaling appears with strange 
regularity in all known experimental tests of Bell's inequality...

\section{Beyond quantum}

Finally, I present another motivation to create  an open access data base. This motivation may be not so important for the quantum community, but it has to be taken into account 
from the viewpoint of  development of science in general. As the organizer of the longest series of conferences on quantum foundations, the V\"axj\"o 
series, I see that the stream of people questioning  QM is not vanishing at all, may be nowadays even more people question QM than  30-40 years ago,
Why? The answer is known to everybody: because, as Einstein, people are not satisfied that the greatest physical theory is foundationally just  an operational formalism 
for calculation of probabilities. Many of these scientists are not ``scientific outsiders'', they are  qualified physicists. They are sometimes 
suspicious that there are no 
open access data, they suspect  that quantum foundational experiments have 
deeper statistical complexity than it is typically claimed in the papers; they want to compare their theories with the real data.

 Of course, experimenters by uploading data to websites can expect that their statistical 
conclusions may be criticized. And they expectations are correct, they would get more critical publications. But this is precisely as science has to work; this may stimulate
experimenters to employ more experts in foundations who will be busy with the critical analysis of the critical publications. We all know that the chance that in future something 
useful would be found beyond quantum is not so high, cf. \cite{Beyond}. However, it is still nonzero. Recently one of the best experimenters working in quantum foundations 
told me: {\small ``Well, the most funny thing would happen if data from the final loophole free Bell experiment were not violate 
Bell's inequality.''}

And besides of the natural scientific unsatisfactoriness by the situation in which  one simply should ``shut up and calculate'', we have not forget about the real black cloud 
at the quantum horizon: the problem of unification of QM and general relativity. Nowadays many people start to suspect that this problem is unsolvable without to
 go beyond, either of QM or general relativity. It seems that by playing with more advanced models of noncommutative mathematics one would not be able to come to the 
``greatest unification.''

But once again: the conventional quantum community can ignore this ``beyond quantum problem''. My main motivation for creation of the open access
data-base for the rough experimental data is to provide a possibility for a detailed statistical analysis of this data by independent quantum researchers.    

The situation such that after about 20 years after Weihs' experiment \cite{Weihs0} (closed the locality loophole) the data from this experiment is the only openly available for the quantum 
community (at least it was) is really unacceptable.

\section*{Acknowledgment}

This note was written after the my recent visit  to NIST and I would like to thank A. Migdall and S. Polyakov for hospitality. This visit stimulated me to write 
finally such a note. I was thinking to do this a few times, but was not sure whether it would be meaningful.  When I found that even in such a place people suffer 
of the absence of good rough data on Bell's test, I decided that something finally has to be done. I also would like to thank  A. Migdall and S. Polyakov, 
Ya. Zhang, S. Glancy, and E. Knill for discussions on technicalities involved in Bell-type experiments and  
S. Ramelow, B. Wittmann, J. Kofler, and R. Ursin for similar discussions during  my visits to IQOQI in 2013, 2014.  I thank G. Adenier and I. Basieva 
for many years of free discussions on the Bell test.

\end{document}